\documentclass[twocolumn, amsmath,nobibnotes,nofootinbib,showpacs,aps,prb]{revtex4}  
\usepackage[dvips]{graphicx,color}
\begin{document}
%\title{Short-range order in octahedral site of inverse spinel structure of NiFe$_2$O$_4$ and occurrence of large ferroelectricity}
\title{Ferroelectric order associated with an ordered occupancy at the octahedral site of the inverse spinel structure of multiferroic NiFe$_2$O$_4$}
\author{J. K. Dey$^1$}
\author{A. Chatterjee$^1$}
\author{S. Majumdar$^1$}
\author{A.-C. Dippel$^2$}
\author{O. Gutowski$^2$}
\author{M. v. Zimmermann$^2$}
\author{S. Giri$^1$}
\email{sspsg2@iacs.res.in} 
\affiliation{$^1$School of Physical Sciences, Indian Association for the Cultivation of Science, Jadavpur, Kolkata 700032, India}
\affiliation{$^2$Deutsches Elektronen-Synchrotron DESY, Notkestr. 85, 22607 Hamburg, Germany}

\begin{abstract}
We report a ferroelectric order at $\sim$ 98 K for NiFe$_2$O$_4$, which carries an inverse spinel structure with a centrosymmetric $Fd\overline{3}m$ structure at room temperature. The value of spontaneous electric polarization is considerably high as $\sim$ 0.29 $\mu$C/cm$^2$ for 5 kV/cm poling field. The electric polarization decreases considerably ($\sim$ 17 \%) around liquid nitrogen temperature upon application of 50 kOe field, proposing a significant magnetoelectric coupling. The synchrotron diffraction studies confirm a structural transition at $\sim$ 98 K  to a noncentrosymmetric structure of $P4_{1}22$ space group. The occurrence of polar order is associated with an ordered occupancy of Ni and Fe atoms at the octahedral sites of the $P4_{1}22$ structure, instead of random occupancies at the octahedral site of the inverse spinel structure. The results propose that NiFe$_2$O$_4$ is a new type-II multiferroic material.

\end{abstract}
\pacs{75.85.+t, 77.80.-e}
\maketitle
%\cite{zhang_bat,zhou_bat,zhang_mag,ferr_bio,jord_bio,zhang,verwey,coey,wright,zuo,senn,kato1,bar}
\section{Introduction}
Nickel ferrite with the chemical formula NiFe$_2$O$_4$ (NFO) attracts special attention for the similar significant physical properties with magnetite. Magnetite has been recognized as an oldest known and the most studied magnet in science with diverse applications in many areas such as  rechargeable batteries,\cite{zhang_bat,zhou_bat} magnetic recording,\cite{zhang_mag} medicine, and biology\cite{ferr_bio,jord_bio} and continues to excite with the complex and intriguing fundamental properties.\cite{zhang,verwey,coey,wright,zuo,senn,kato1,bar} Despite of many similarities, investigations on NFO have been less attempted and the unfolded issues need to be explored.

\begin{figure}[t]
\centering
\includegraphics[width = \columnwidth]{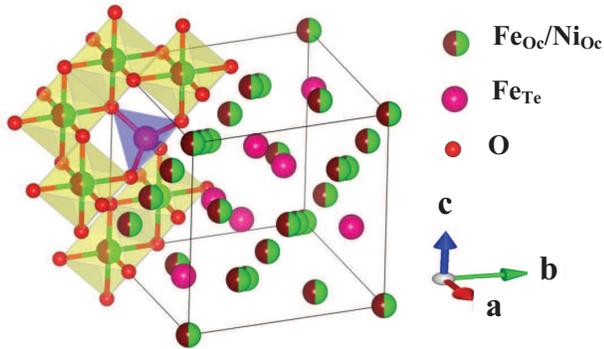}
\caption {(Color online) Atomic positions in the inverse spinel structure of NFO. A portion of connecting (Fe,Ni)O$_6$ octahedra around a FeO$_4$ tetrahedron is also depicted, where "Oc" and "Te" in the suffix indicate the octahedron and tetrahedron.} 
\label{str}
\end{figure}

Magnetite crystallizes in the inverse spinel structure ($Fd\overline{3}m$) at room temperature. The formal chemical formula of Fe$_3$O$_4$ can be represented as Fe$^{3+}_{\rm Te}$[Fe$^{2+}$Fe$^{3+}$]$_{\rm Oc}$O$^{2-}_4$, where one third of Fe occupies the tetrahedral (Te) site in Fe$^{3+}$ state, and the rest Fe$^{2+}$ and Fe$^{3+}$ of equal number occupy the octahedral (Oc) site. Analogous to Fe$_3$O$_4$, NFO also carries the inverse spinel structure, where Ni$^{2+}$ replaces the Fe$^{2+}$ at the octahedral site in the $Fd\overline{3}m$ structure. Inverse spinel structure of NFO is illustrated in Fig. \ref{str}, where tetrahedral (8$b$) site is occupied by Fe$^{3+}$ and octahedral (16$c$) site is occupied by Ni$^{2+}$ and Fe$^{3+}$ with the 50 \% occupancy, as highlighted by the bi-color atoms in the figure. Figure also partially depicts that octahedra are connected to each other and linkage of the octahedra with a tetrahedral unit through the O atoms. Similar to the ferrimagnetic order of magnetite at high temperature (860 K),\cite{kato1} NFO also orders ferrimagnetically at 853 K.\cite{chika} Low temperature M\"{o}ssbauer study in external magnetic field proposed a collinear magnetic structure of NFO.\cite{chap} The neutron diffraction study also confirms the collinear ferrimagnetic order at room temperature.\cite{ugen,hast} Plenty of works were performed on the doping effect of NFO, which significantly influenced the magnetic, electronic, and structural properties.\cite{chap,bhar1,bhar2,rez1,rez2,rez3,pari,ugen1,shan,saha} Nanoscale properties of NFO have been investigated providing rich consequences in the magnetic and dielectric properties for the nanoparticles\cite{wang,rahi,mene,wang1,sep,ush,cve} and films.\cite{fri,riga,lud} The spin-filtering effect has been proposed using the first principles density functional theory for NFO,\cite{szo,caff} which was corroborated for NFO ultrathin films.\cite{mat} Raman studies were carried out on NFO and argued possible occurrence of the short range order at the octahedral site either in the tetragonal $P4_{1}22$ or $P4_{3}22$ symmetry at low temperature.\cite{ili,iva} The proposed results are analogous to the symmetry lowering at the Verwey temperature for magnetite. Unlike adequate structural investigations on magnetite,\cite{verwey,coey,wright,zuo,senn} the low temperature structural studies of NFO are still lacking, which need to be probed. 

Herein, we report the synchrotron diffraction studies over a wide temperature range of 10$-$300 K and the analysis of the diffraction data using Rietveld refinement confirms a structural transition to a noncentrosymmetric structure ($P4_{1}22$) at $\sim$ 98 K. The results further suggest that the random occupancies at the octahedral site of inverse spinel structure no longer exist below $\sim$ 98 K. Instead, an ordered occupancy of Ni and Fe is proposed. Intriguingly, the structural transition to a noncentrosymmetric structure is associated with the occurrence of a spontaneous ferroelectric (FE) order with the FE $T_C$ close to 98 K. The value of electric polarization ($P$) is considerably high as $\sim$ 0.29 $\mu$C/cm$^2$ for a 5 kV/cm poling field, which is comparable to those values of promising type-II multiferroics.\cite{khom,dey_FE1,jhuma,indra_FE,kimura} Till date, ferroelectric order has been reported for a vanadate,\cite{zha_vanadate} few chromate oxides,\cite{dey_FE2,singh,yama,choi,yang,mai,bush} few chromate sulphides,\cite{dey_FE3,hem1,cata,web,lin} and a cobalt aluminate oxide,\cite{ghara} which had an ordered spinel structure in the paraelectric state. However, the multiferroic order has been rarely explored in a system involving inverse spinel structure, except for the classic example of ferroelectric order in magnetite.\cite{kato,miya,med,ale,taka,sch,zie,dey_APL,mis} Current results are analogous to the observed FE order at low temperature for magnetite, although origin of FE order remains unsettled.\cite{kato,miya,med,ale,taka,sch,zie,dey_APL,mis} Here, possible origin of FE order in NFO is discussed based on the structural analysis.   

%%%%%%%%%%%%%%%%%%%%%%%%%%%%%%%%%%%%%%%%%%%%%%%%%%%%%%%%%%%%%%%%%%%%%%%%%%%%%%%%%%%%%%%%%%%%%%%%%%%%%%%%%%%%%%%%%%%%%%%
%%%%%%%%%%%%%%%%%%%%%%%%%%%%%%%%%%%%%%%%%%%%%%%%%%%%%%%%%%%%%%%%%%%%%%%%%%%%%%%%%%%%%%%%%%%%%%%%%%%%%%%%%%%%%%%%%%%%%%%

\section{Experimental details}
Nickel ferrite with formula NiFe$_2$O$_4$ is prepared using solid state reaction.\cite{chap} The chemical composition is checked by the x-ray diffraction studies at room temperature recorded in a PANalytical x-ray diffractometer (Model: X' Pert PRO) using the Cu K$\alpha$ radiation. The single phase chemical composition is further confirmed by the synchrotron x-ray diffraction studies recorded with a wavelength of 0.1422 \AA~ (87.1 keV) at the P07 beamline of PETRA III, Hamburg, Germany in the temperature range of 10-300 K. The synchrotron  diffraction data are analyzed using the Rietveld refinement with a commercially available MAUD software. X-ray photoemission spectroscopy (XPS) is recorded in a spectrometer of Omicron Nanotechnology. The powder sample pressed into a pellet is used for the dielectric measurements using an E4980A LCR meter (Agilent Technologies, USA) equipped with a commercial PPMS evercool-II system of Quantum Design. The pyroelectric current ($I_p$) is recorded in an electrometer (Keithley, model 6517B) at a constant temperature sweep rate. In all the measurements the electrical contacts are fabricated using an air drying silver paint. The poling electric fields are applied during cooling processes and $I_p$ measurements are carried out in the warming mode in absence of electric field. Before the measurement of $I_p$, the electrical connections are short circuited and waited for a sufficiently long time. Magnetization is measured in a commercial magnetometer of Quantum Design (MPMS, evercool).

%%%%%%%%%%%%%%%%%%%%%%%%%%%%%%%%%%%%%%%%%%%%%%%%%%%%%%%%%%%%%%%%%%%%%%%%%%%%%%%%%%%%%%%%%%%%%%%%%%%%%%%%%%%%%%%%%%%%%%
\begin{figure}[t]
\centering
\includegraphics[width = \columnwidth]{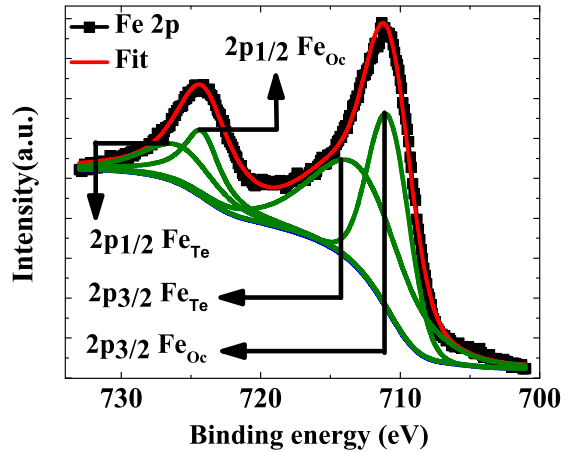}
\caption {(Color online) Fe 2$p$ core level XPS of NFO. Deconvolution of the curves along with the fit are shown by the solid curves.} 
\label{xps}
\end{figure}
%%%%%%%%%%%%%%%%%%%%%%%%%%%%%%%%%%%%%%%%%%%%%%%%%%%%%%%%%%%%%%%%%%%%%%%%%%%%%%%%%%%%%%%%%%%%%%%%%%%%%%%%%%%%%%%%%%%%%%

%%%%%%%%%%%%%%%%%%%%%%%%%%%%%%%%%%%%%%%%%%%%%%%%%%%%%%%%%%%%%%%%%%%%%%%%%%%%%%%%%%%%%%%%%%%%%%%%%%%%%%%%%%%%%%%%%%%%%%
\begin{figure*}[t]
\centering
\includegraphics[width = 1.7\columnwidth]{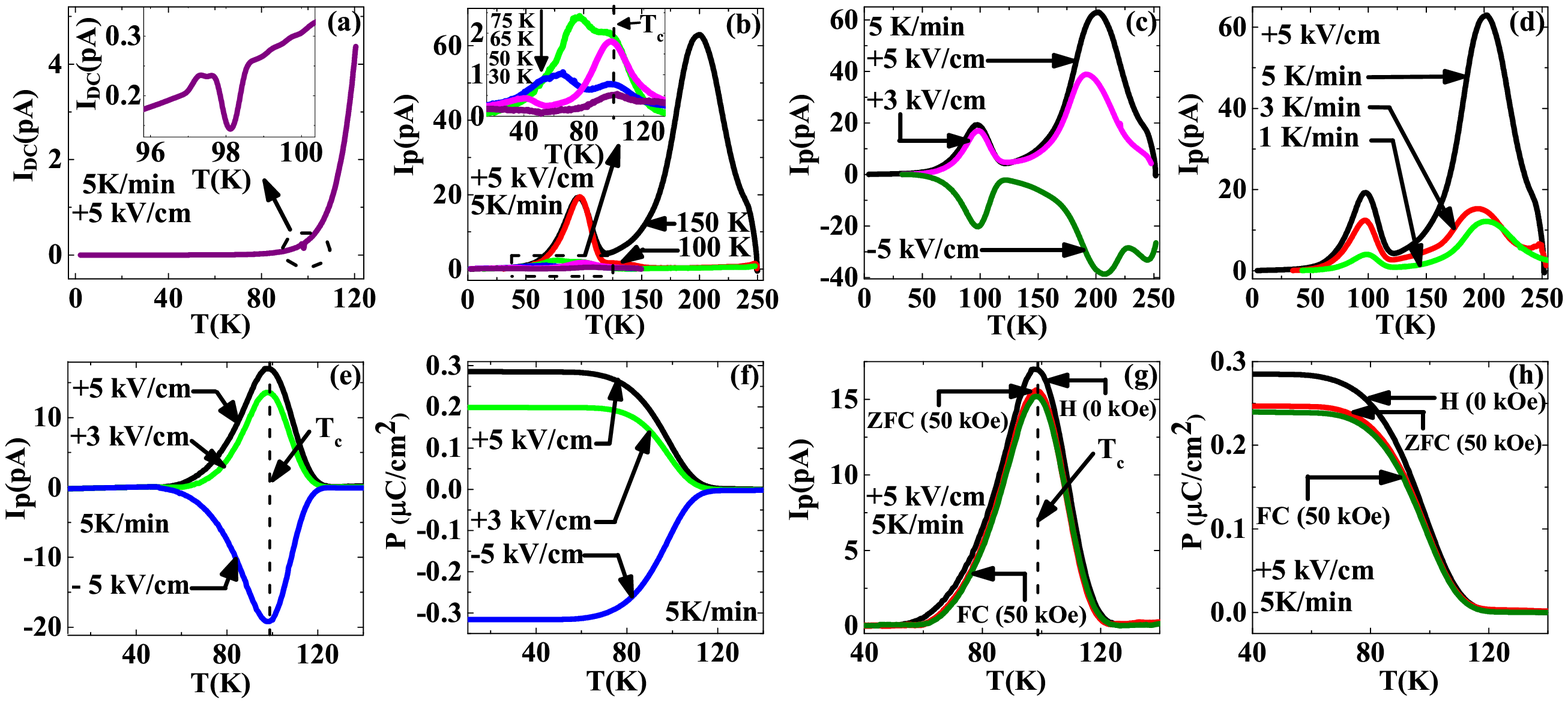}
\caption {(Color online) (a) Temperature ($T$) variation of DC-biased current ($I_{DC}$) with a 5 kV/cm bias field. Inset highlights the "dip" in $I_{DC}(T)$ at 98 K. $T$ variations of (b) pyroelectric current ($I_P$) for different poling temperatures ($T_{\rm pole}$) at poling field ($E$) of 5 kV/cm, (c) $I_P$ for different $E$ and $T_{\rm pole}$ = 150 K, (d) $I_P$ for different temperature sweep rates and $E$ = 5 kV/cm, (e) $I_P$ for different $E$ and $T_{\rm pole}$ = 150 K after subtracting the extrinsic thermally stimulated depolarization currents, (f) electric polarization ($P$) at different $E$, (g) $I_P$, and (h) $P$ at $E$ = 5 kV/cm in zero magnetic field and 50 kOe using both the ZFC and FC modes. Inset of (b) highlights the peaks at FE transition ($T_C$).} 
\label{pyro}
\end{figure*}
%%%%%%%%%%%%%%%%%%%%%%%%%%%%%%%%%%%%%%%%%%%%%%%%%%%%%%%%%%%%%%%%%%%%%%%%%%%%%%%%%%%%%%%%%%%%%%%%%%%%%%%%%%%%%%%%%%%%%%

\section{Experimental results and discussions}
%\subsection{Spontaneous electric and magnetic polarizations}
 The XPS measurements are carried out in order to detect possible existence of Fe$^{2+}$ attributed to the oxygen non-stoichiometry, \cite{jaff} or involving the doping effect.\cite{ugen} Figure \ref{xps} depicts the Fe 2$p$ core level spectrum of NFO, where solid curve on the experimental data shows the satisfactory fit. The deconvolution of the peak into two components of 50 \% area under the curves correspond to the octahedral 2$p_{3/2}$ and 2$p_{1/2}$ peaks, and the tetrahedral 2$p_{3/2}$ and 2$p_{1/2}$ peaks. The peaks are observed at 712.9 and 710.8 eV for tetrahedral Fe 2$p_{3/2}$ and octahedral Fe 2$p_{3/2}$, respectively, and at 725.8 and 724.2 eV for tetrahedral Fe 2$p_{1/2}$ and octahedral Fe 2$p_{1/2}$, respectively. The results are consistent with the literature.\cite{jaff,huf} The fit of the XPS results indicate the absence of Fe$^{2+}$ component and confirms the oxygen stoichiometric compound. 

In order to observe possible occurrence of ferroelectric order in NFO, we first measure DC-bias current ($I_{DC}$) using the bias electric field (BE) method, as recently proposed to identify genuine occurrence of ferroelectricity.\cite{cde,terada} In BE method we record $I_{DC}$ with a bias electric field of 5 kV/cm in the warming mode for a heating rate of 5 K/min after cooling the sample down to 2 K in zero electric field. The results of BE measurements in temperature range of 2$-$120 K are depicted in Fig. \ref{pyro}(a). The inset of Fig. \ref{pyro}(a) highlights the close view around 98 K, at which a clear signature of a 'dip' is observed. In order to further confirm it, the pyroelectric current is recorded at different poling temperatures ($T_{\rm pole}$) for a +5 kV/cm poling field. Here, sample is always cooled from the selected $T_{\rm pole}$ down to a low temperature (minimum up to 2 K) and $I_p$ is measured during warming mode in zero field. The results of $I_p$ with $T$ at different $T_{\rm pole}$ are depicted in Fig. \ref{pyro}(b) for a heating rate of 5 K/min. Here, different values of $T_{\rm pole}$ are selected as  the representative of poling temperatures below, above, and close to $\sim$ 98 K, at which a clear signature of a 'dip' is observed in the BE method. In all the cases an apparent peak is always observed around $\sim$ 98 K, as also depicted in the inset of the figure. The peak height around 98 K is nearly same for $T_{\rm pole} \geq$ 98 K (at 100 and 150 K). The peak height decreases considerably for $T_{\rm pole} \leq$ 98 K at 75, 65, 50, and 30 K, which is further highlighted in the inset of Fig. \ref{pyro}(b). 

%%%%%%%%%%%%%%%%%%%%%%%%%%%%%%%%%%%%%%%%%%%%%%%%%%%%%%%%%%%%%%%%%%%%%%%%%%%%%%%%%%%%%%%%%%%%%%%%%%%%%%%%%%%%%%%%%%%%%%%
%\begin{figure}[b]
%\centering
%\includegraphics[width = \columnwidth]{Fig_NFO_4.eps}
%\caption {(Color online) Thermal variations of pyroelectric current ($I_P$) at different poling fields for (a) NiFe$_{1.98}$Yb$_{0.02}$O$_4$ and (c) Ni$_{1.98}$Zn$_{0.02}$Fe$_2$O$_4$, and the corresponding electric polarizations ($P$) in (b) and (d), respectively.} 
%\label{MD}
%\end{figure}
%%%%%%%%%%%%%%%%%%%%%%%%%%%%%%%%%%%%%%%%%%%%%%%%%%%%%%%%%%%%%%%%%%%%%%%%%%%%%%%%%%%%%%%%%%%%%%%%%%%%%%%%%%%%%%%%%%%%%%%

We further note that another peak is observed close to $T_{\rm pole}$, as depicted in the inset of the figure for $T_{\rm pole}$ below $\sim$ 98 K. This peak does not engage with the phase transition at 98 K, in accordance with the previous reports.\cite{terada,ngo} In addition to these peaks, another very large peak is observed around $\sim$ 200 K, which occurs for $T_{\rm pole}$ at 150 K. The occurrence of a very large peak is rather consistent with those reported by Ngo {\it et al}.\cite{ngo} In the current investigation, the large peak around 200 K always appears for the measurement at different poling fields and heating rates, when $T_{\rm pole}$ is considered as 150 K, as depicted in Figs. \ref{pyro}(c) and \ref{pyro}(d), respectively. This large peak reverses for negative poling field, as depicted in Fig. \ref{pyro}(c). Importantly, this large peak disappears when $T_{\rm pole}$ is $\leq$ $\sim$ 98 K, as shown in Fig. \ref{pyro}(b). In fact, this peak disappears for $T_{\rm pole}$ = 100 K, as depicted in Fig. \ref{pyro}(b) and confirm that the high-temperature large peak appears due to the extrinsic thermally stimulated depolarization currents (TSDC).\cite{terada,ngo,chen,liu} Thus evident signature of a peak around $\sim$ 98 K in both the pyroelectric current measurement and BE measurement confirms genuine occurrence of the ferroelectricity with a FE $T_C$ at 98 K. 
    
Figure \ref{pyro}(e) depicts $I_p$ with $T$ for different poling fields ($E$) and $T_{\rm pole}$ = 150 K after subtracting the high temperature TSDC components. The large TSDC component is subtracted using method suggested by Ngo {\it et al}.\cite{ngo,chen} After subtracting the TSDC component the $I_p$($T$)s recorded at different heating rates are integrated over time, which nearly reproduce the same $P-T$ curve, pointing that the detrapped charges, if they exist, do not contribute appreciably to the $I_p$-values. Time-integrated $I_p$ providing $P$ as a function of $T$ is depicted in Fig. \ref{pyro}(f) for different values of $E$. Reversal of $P$ due to a change in sign of $E$ further signifies the ferroelectric behavior of NFO. The saturated value of $P$ in the current investigation is $\sim$ 0.29 $\mu C$/cm$^2$ for $E$ = 5 kV/cm, which is significantly large compared to the values of the promising type-II multiferroics.\cite{khom,dey_FE1,jhuma,indra_FE,kimura} 

In order to probe possible magnetoelectric (ME) coupling, the effect of magnetic field ($H$) on $P(T)$ is investigated. The $I_p(T)$ is recorded with $H$ in both the zero-field cooled (ZFC) and field-cooled (FC) mode.\cite{indra_JPCM} In case of ZFC mode sample is cooled down to $\sim$ 10 K in zero-field and a magnetic field of 50 kOe is applied in the warming mode during measurement of $I_p$. For FC mode sample is cooled in field of 50 kOe and $I_p$ is recorded during warming in zero magnetic field. The results are summarized in Figs. \ref{pyro}(g) and \ref{pyro}(h). The results indicate absence of noticeable change for the measurements in ZFC and FC modes. However, a definite change is observed for the measurement in $H$ = 50 kOe. We note that a significant decrease of $P$ is observed below FE $T_C$ due to $H$, as also shown in the figure, which indicates existence of a magnetoelectric (ME) coupling. The percentage of decrease of $P$ is considerable as $\sim$ 17 \% close to liquid nitrogen temperature.

%%%%%%%%%%%%%%%%%%%%%%%%%%%%%%%%%%%%%%%%%%%%%%%%%%%%%%%%%%%%%%%%%%%%%%%%%%%%%%%%%%%%%%%%%%%%%%%%%%%%%%%%%%%%%%%%%%%%%%%
\begin{figure}[t]
\centering
\includegraphics[width = \columnwidth]{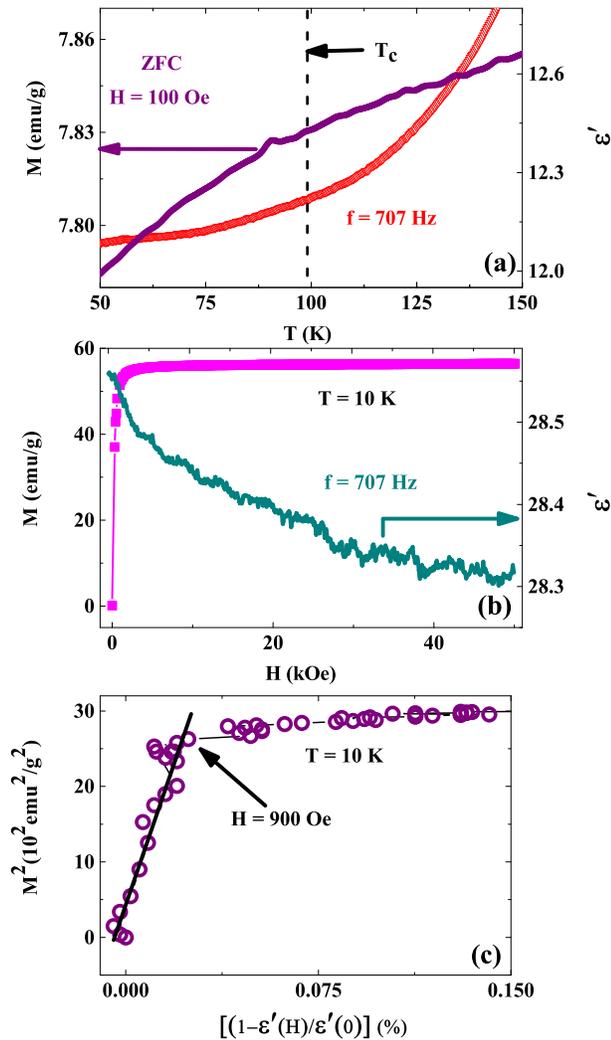}
\caption {(Color online) Thermal variations of (a) ZFC magnetization ($M$) in left axis and real part ($\epsilon^{\prime}$) of dielectric permittivity at frequency, $f$ = 707 Hz in right axis around FE $T_C$. Magnetic field ($H$) dependence of (b) $M$ (left axis) and $\epsilon^{\prime}$ (right axis) at 10 K. (c) Plot of $M^2$ with [1-$\epsilon^{\prime}(H)/\epsilon^{\prime}$(0)](\%) at 10 K. Straight line depicts the linear fit below 900 Oe.} 
\label{MD}
\end{figure}
%%%%%%%%%%%%%%%%%%%%%%%%%%%%%%%%%%%%%%%%%%%%%%%%%%%%%%%%%%%%%%%%%%%%%%%%%%%%%%%%%%%%%%%%%%%%%%%%%%%%%%%%%%%%%%%%%%%%%%%

Magnetization ($M$) recorded in ZFC mode and dielectric permittivity ($\epsilon$) at $H$=0 are measured with $T$. The values of $M$ and real component ($\epsilon^{\prime}$) of $\epsilon$ with $T$ are depicted around FE $T_C$ in Fig. \ref{MD}(a), where a vertical broken line shows the position of FE $T_C$. In both the cases of $M(T)$ and $\epsilon^{\prime}(T)$, we could not detect any noticeable signature of FE $T_C$. Absence of the convincing signature in $\epsilon^{\prime}(T)$ close to FE $T_C$ has also been observed in other multiferroics.\cite{indra_FE,indra_JPCM,maher_dielec,ghosh1_dielec,ghosh2_dielec,raj_dielec,indra_2018} The overlapping of the intrinsic component with the extrinsic contributions in $\epsilon$ such as the grain boundary and the sample-electrode interface effects may lead to the weak signature or absence of any signature around FE $T_C$. Figure \ref{MD}(b) depicts the magnetization curve, displaying a soft ferromagnetic character and variation $\epsilon^{\prime}$ with $H$ at 10 K. At 10 K a small decrease of $\epsilon^{\prime}$ is observed with increasing $H$. The percentage of magnetodielectric response, defined as [1 - $\epsilon^{\prime}(H)/\epsilon^{\prime}(H = 0)]\times 100$, is estimated to be $\sim$ 0.9 \% at 10 K for $H$ = 50 kOe, which is comparable to the results for different multiferroics such as CoCr$_2$S$_4$,\cite{dey_FE3} MCr$_2$O$_4$ (M = Mn, Co, Ni),\cite{spark,mufti} BiMnO$_3$,\cite{kimu} ZnCr$_2$O$_4$,\cite{dey} Sm$_2$BaNiO$_5$,\cite{indra_FE} and $R$CrO$_4$.\cite{indra_2018} 
The magnetodielectric response is phenomenologically described by the Ginzburg-Landau theory for a second-order phase transition and is attributed to a ME coupling term $\gamma P^{2}M^{2}$ in the thermodynamic potential given by
\begin{equation}
%\Phi = \Phi_0 + \alpha P^2 + \frac{\beta}{2} P^4 - PE + \alpha^{\prime} M^2 + \frac{\beta^{\prime}}{2} M^4 - MH + \gamma P^2 M^2.
\Phi = \Phi_0 + \alpha P^2 - PE + \alpha^{\prime} M^2 - MH + \gamma P^2 M^2.
\end{equation}
The $\alpha$, $\alpha^{\prime}$, and $\gamma$ in the above equation are the constants and functions of temperature. The linear dependence between squared magnetization and magnetodielectric response in the low field regime indicate that the magnetodielectric coupling term $\gamma P^2 M^2$ of the Ginzburg-Landau theory is significant. Here, linear plot of $M^2$ with [1-$\epsilon^{\prime}(H)/\epsilon^{\prime}(H = 0)]\times 100$ is observed below $\sim$ 0.9 kOe, as indicated by an arrow in Fig. \ref{MD}(b), which has also been reported for the multiferroics with spinel structure \cite{dey_FE3,spark,mufti,dey} and other multiferroic oxides.\cite{kimu,indra_FE,indra_2018}

%*************************************************************************************************
\begin{table}[b] 
%\centering
\caption{Atomic positions of NFO with $Fd\overline{3}m$ (No. 227, Z=8) and $P4_{1}22$ (No. 91, Z=4) symmetries at 300 and 90 K, respectively.}
\label{table}
\begin{center}
\begin{tabular}{ccccccc}		%cccccc
\hline
\hline
$T$ & atoms & $x$ & $y$ & $z$ & occupancy & site \\
\hline
300 K & Ni$^{2+}$/Fe$^{3+}$ 	&	 0.00000  &  0.00000  &  0.00000  &  1.0 (50 \%)   &    16$c$ \\
%			& Fe$^{3+}$ 	&	 0.00000  &  0.00000  &  0.00000  &  0.5   &    16$c$ \\
			& Fe$^{3+}$ 	&  0.62500  &  0.62500  &  0.62500  &  1.0   &    8$b$  \\
			& O$^{2-}$    &  0.24797  &  0.24797  &  0.24797  &  1.0   &    32$e$ \\
\hline
90 K  & Ni$^{2+}$ 	&	 0.00000  &  0.75000  &  0.00000  &  0.5   &    4$a$ \\
			& Fe$^{3+}$ 	&	 0.50000  &  0.75000  &  0.00000  &  0.5   &    4$b$ \\
			& Fe$^{3+}$ 	&  0.75000  &  0.75000  &  0.37500  &  1.0   &    4$c$  \\
			& O1$^{2-}$   &  0.50792  &  0.75000  &  0.25000  &  1.0   &    8$d$ \\
			& O2$^{2-}$   &  0.03935  &  0.25000  &  0.75000  &  1.0   &    8$d$ \\
\hline   
\hline
\end{tabular}
\end{center}
\end{table}      

%%%%%%%%%%%%%%%%%%%%%%%%%%%%%%%%%%%%%%%%%%%%%%%%%%%%%%%%%%%%%%%%%%%%%%%%%%%%%%%%%%%%%%%%%%%%%%%%%%%%%%%%%%%%%%%%%%%%%%%

\begin{figure}[t]
\centering
\includegraphics[width = \columnwidth]{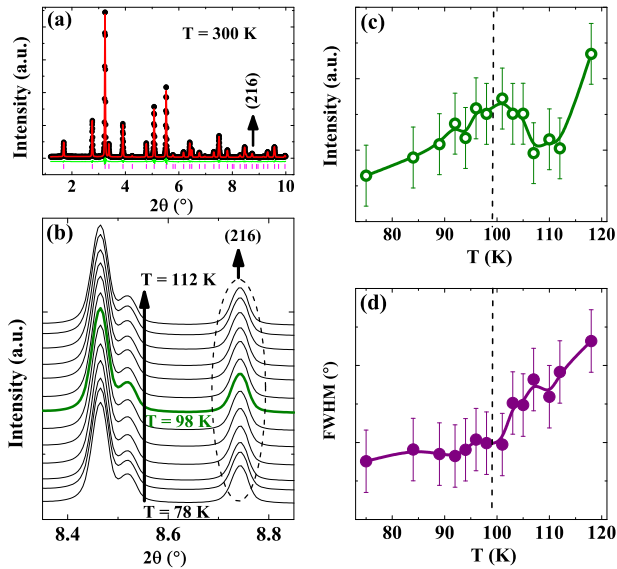}
\caption {(Color online) Synchrotron diffraction patterns at (a) 300 K for NFO. Solid curve represents the fit using Rietveld refinement. (b) Selected diffraction peaks at different temperatures around FE $T_C$. $T$ dependence of (c) integrated intensities and (d) FWHM of the (216) peak.} 
\label{XRD-RT}
\end{figure}

NiFe$_2$O$_4$ crystallizes in the inverse spinel structure with a centrosymmetric $Fd\overline{3}m$ (No. 227, Z=8) space group at 300 K. At 300 K a satisfactory fit of the diffraction pattern using Rietveld refinement is depicted by the solid curve in Fig. \ref{XRD-RT}(a). The refined positional coordinates are given in Table I considering origin at the octahedral position. The reliability parameters, $R_w$(\%) $\sim$ 4.98, $R_{exp}$(\%) $\sim$ 3.04, and $\chi^2$(\%) $\sim$ 1.76 at 300 K are reasonable. In order to search possible structural change associated with the occurrence of polar order, the selected diffraction peaks are plotted with temperature around FE $T_C$, as depicted in Fig. \ref{XRD-RT}(b). The small changes around FE $T_C$ are noted, which are evident in the thermal variations of intensity and full width at half maxima (FWHM) of the (216) diffraction peak, as depicted in Figs. \ref{XRD-RT}(c) and \ref{XRD-RT}(d), respectively. In Fig. \ref{XRD-RT}(c) an apparent signature with a maximum is observed at FE $T_C$. This signature at FE $T_C$ may be associated with the change in the scattering cross section. Hence a change in the scattering amplitude may be correlated to this intensity change and indicates possible occurrence of a structural transition. The change in intensity is similar to that observed for the reported ferroelectric materials.\cite{indra_FE,dey_FE2,dey_FE3,indra_2018} As depicted in Fig. \ref{XRD-RT}(d), an anomaly in FWHM is also noted around FE $T_C$, where the decreasing trend in the thermal variation of FWHM becomes flat below $\sim$ FE $T_C$.

%%%%%%%%%%%%%%%%%%%%%%%%%%%%%%%%%%%%%%%%%%%%%%%%%%%%%%%%%%%%%%%%%%%%%%%%%%%%%%%%%%%%%%%%%%%%%%%%%%%%%%%%%%%%%%%%%%%%%%%
%%%%%%%%%%%%%%%%%%%%%%%%%%%%%%%%%%%%%%%%%%%%%%%%%%%%%%%%%%%%%%%%%%%%%%%%%%%%%%%%%%%%%%%%%%%%%%%%%%%%%%%%%%%%%%%%%%%%%%%

\begin{figure}[t]
\centering
\includegraphics[width = \columnwidth]{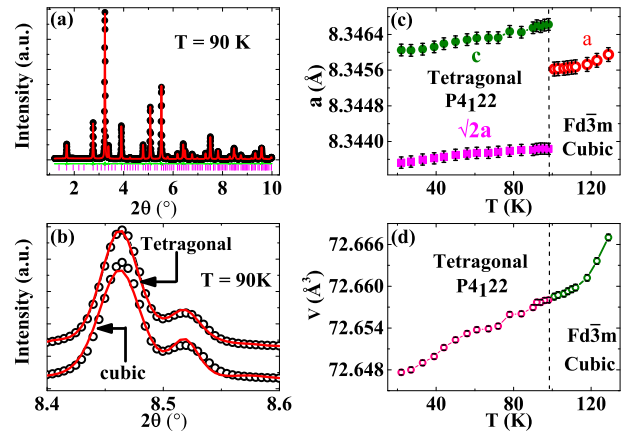}
\caption {(Color online) Synchrotron diffraction patterns at (a) 90 K for NFO. Solid curve represents the fit using Rietveld refinement. Rietveld refinements are compared with (b) tetragonal $P4_{1}22$ and cubic $Fd\overline{3}m$ space groups at 90 K. $T$ dependence of (c) lattice constants and (d) unit cell volume ($V$).} 
\label{XRD}
\end{figure}

Till date, the low temperature structural studies have not been attempted for NFO. Recently, first principles calculations proposed possible symmetry lowering to either noncentrosymmetric $P4_{1}22$ or centrosymmetric $Imma$ space group at low temperature for NFO.\cite{fri,caff,frit,jong} The Raman spectroscopy in epitaxial films as well as single crystal of NFO also proposed possible B site 1:1 ordering in the tetragonal $P4_{1}22$ symmetry.\cite{ili,iva} The available neutron diffraction studies down to the room temperature confirmed a collinear ferrimagnetic order.\cite{ugen,hast} If we consider the collinear ferrimagnetic order at low temperature around FE $T_C$, a structural transition to a non-centrosymmetric structure is justified to interpret occurrence of the ferroelectricity in NFO. Thus we incorporate AMPLIMODE\cite{oro} and ISODISTORT\cite{camp} soft wares to find out the possible non-centrosymmetric space groups below FE $T_C$. We note that $P4_{1}22$ space group is one of the possible candidates involving structural transition and consistent with that proposed from the first principles calculations\cite{fri,caff,frit,jong} as well as Raman studies.\cite{ili,iva} A satisfactory refinement of the diffraction pattern at 90 K using tetragonal $P4_{1}22$ space group is depicted by the solid curve in Fig. \ref{XRD}(a). The refined coordinates of the atoms are listed in Table I. The values of the reliability parameters, $R_w$(\%) $\sim$ 5.71, $R_{exp}$(\%) $\sim$ 3.04, and $\chi^2$(\%) $\sim$ 1.87 at 90 K are satisfactory. An example of the comparative fit of the selected diffraction peaks is highlighted in Fig. \ref{XRD}(b) using both the high temperature $Fd\overline{3}m$ structure and the proposed $P4_{1}22$ space group. The results clearly demonstrate the better fit of the diffraction peaks with the tetragonal $P4_{1}22$ space group and proposes a structural transition to the tetragonal $P4_{1}22$ structure from the cubic $Fd\overline{3}m$ structure.

%%%%%%%%%%%%%%%%%%%%%%%%%%%%%%%%%%%%%%%%%%%%%%%%%%%%%%%%%%%%%%%%%%%%%%%%%%%%%%%%%%%%%%%%%%%%%%%%%%%%%%%%%%%%%%%%%%%%%%%
%%%%%%%%%%%%%%%%%%%%%%%%%%%%%%%%%%%%%%%%%%%%%%%%%%%%%%%%%%%%%%%%%%%%%%%%%%%%%%%%%%%%%%%%%%%%%%%%%%%%%%%%%%%%%%%%%%%%%%%

Thermal ($T$) variations of the lattice parameters are obtained from the refinement of the diffraction patterns at different temperatures. The $T$ variations of lattice constants are depicted in Figs. \ref{XRD}(e). Here, the FE $T_C$ in the figure is pointed out by the vertical broken line. A discontinuous change of lattice constants is observed at the FE $T_C$. The $\sqrt{2}a$ and $c$ below the structural transition show a slow decrease with decreasing $T$.   
The results demonstrate that the ferroelectricity in NFO is correlated to this structural transition from the centrosymmetric $Fd\overline{3}m$ to a noncentrosymmetric $P4_{1}22$ structure. The unit cell volume ($V$), as obtained from the lattice constants, is depicted in Fig. \ref{XRD}(f). A change of slope in $V$ is depicted at FE $T_C$.

\begin{figure}[t]
\centering
\includegraphics[width = \columnwidth]{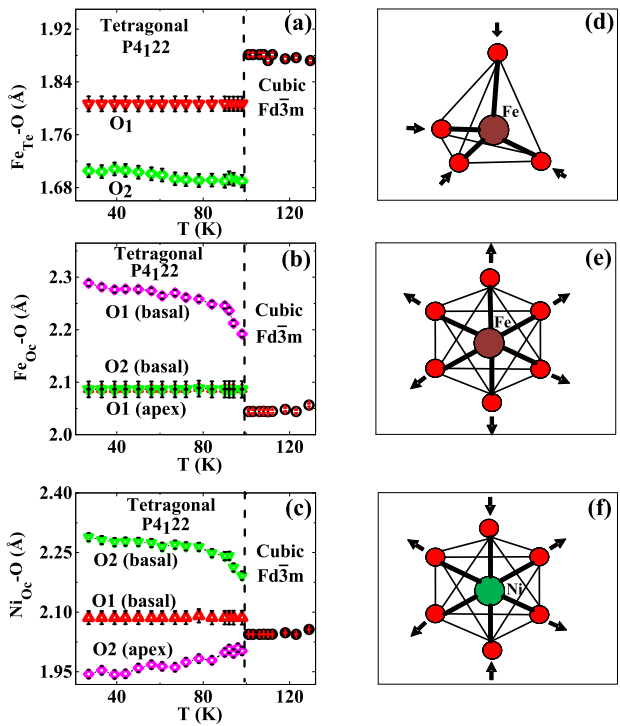}
\caption {(Color online) Thermal variations of the bond lengths of (a) Fe tetrahedra (Fe$_{\rm Te}-$O), (b) Fe octahedra (Fe$_{\rm Oc}-$O), and (c) Ni octahedra (Ni$_{\rm Oc}-$O). Schematic representations of the distortions of the (d) Fe tetrahedron, (e) Fe octahedron, and (f) Ni octahedron.} 
\label{XRD_para}
\end{figure}

A significant structural distortion is noted at the structural transition. In order to understand these distortions microscopically, the bond lengths around Fe and Ni atoms are estimated further. As given in Table I, the $P4_{1}22$ structure carries the two oxygen atoms, such as, O1 and O2. Thus single Fe$-$O bond length in the $Fd\overline{3}m$ structure is converted to two bond lengths corresponding to O1 and O2, respectively due to structural transition. In the $P4_{1}22$ structure Fe atoms occupy the tetrahedral (4$c$) and octahedral (4$b$) sites. Four Fe$-$O bond lengths are possible in the tetrahedral site. Thermal variations of two representative bond lengths corresponding to O1 and O2 are depicted in Fig. \ref{XRD_para}(a). The other two bond lengths behave similarly, which are not shown here. The magnitude of bond lengths decrease considerably at the structural transition, which decrease further with further decreasing temperature and become temperature independent below the $T$-range of $\sim$ 70$-$80 K. The decrease of bond lengths involving O1 and O2 are $\sim$ 4 and $\sim$ 10 \%, respectively at $\sim$ 77 K compared to the value in the paraelectric state. A schematic representation of the distortion in the tetrahedral unit at the structural transition is shown in Fig. \ref{XRD_para}(d), as indicated by the arrows. Thus a contraction of the tetrahedral unit involves the structural transition. 

%*********************************************************************************************************************
%***********************************************************************************************************************

The oxygen atoms involving the bond length along crystallographic $c$ axis are designated as the apex oxygen atoms and the rest four are the basal oxygen atoms in the octahedral structure. Thermal variations of the Fe$-$O bond lengths involving basal and apex oxygen atoms are depicted in Fig. \ref{XRD_para}(b), when Fe atoms occupy the octahedral site. Here, the representative three bond lengths are shown in the figure. The rest three bond lengths behave similarly, which are not shown here. At the structural transition bond lengths increase discontinuously. With further decreasing temperature bond length with apex oxygen remains nearly temperature independent, whereas bond lengths with basal oxygen atoms increase further and become temperature independent below the $T$-range of $\sim$ 70$-$80 K. The increase of bond lengths at $\sim$ 77 K are $\sim$ 2, $\sim$ 2, and $\sim$ 10.5 \% with apex O1, basal O2, and basal O1 atoms, respectively. The schematic representation of the distortion of the octahedral unit at the structural transition is depicted in Fig. \ref{XRD_para}(e). Here, the arrows represent directions of the distortion at the structural transition and propose expansion of the Fe octahedral unit. Similar to the results of Fe octahedral unit, the results of Ni octahedra are shown in Figs. \ref{XRD_para}(c) and \ref{XRD_para}(f). At the structural transition the value of Ni$-$O bond length with apex O2 decreases, whereas the magnitude of the rest two bond lengths involving basal oxygen atoms increases. The results propose a mixed distortion in the Ni octahedra unlike Fe octahedra. A contraction along the apex oxygen atoms and expansions of the octahedra with the basal oxygen atoms are noted, as depicted in Fig. \ref{XRD_para}(f). The decrease of bond lengths with apex O2 is $\sim$ 2 \% and increase of bond lengths with basal O1 and O2 are $\sim$ 2 and $\sim$ 11 \%, respectively.   

The discover of Verwey transition \cite{verwey} in magnetite opened up many unexplored issues.\cite{zhang_bat,zhou_bat,zhang_mag,ferr_bio,jord_bio,zhang,verwey,coey,wright,zuo,senn,kato1,bar} Possible charge ordering of Fe$^{2+}$ and Fe$^{3+}$ at the octahedral site, nature of structural transition, occurrence of ferroelectricity, etc. are few of those intricate issues, which still continue to attract fundamental interests. Analogous to the Verwey transition in magnetite, occurrence of ferroelectricity in NFO driven by the structural transition to the non-centrosymmetric structure also attracts the community for the fundamental interests. The results leave some unsolved issues, which need to be explored both theoretically and experimentally using sophisticated experimental tools. Although the preliminary ZFC magnetization as well as dielectricity does not indicate any convincing signature of ferroelectric order or structural transition, a reasonably strong magnetoelectric coupling is manifested by a $\sim$ 17 \% decrease of polarization upon application of 50 kOe magnetic field close to liquid nitrogen temperature. Thus a delicate issue of possible modulation of the magnetic structure around ferroelectric order needs to be examined by incorporating the structural transition. Recently, Jong {\it et al}. proposed a ferroelectricity in NFO driven by the {\it p-d} hybridization from the first-principles study, where a much larger value of electric polarization of 23 $\mu$C/cm$^2$ along the $z$-direction was estimated.\cite{jong} In the literature the hybridization between the 3$d$  states of the Fe$^{3+}$ cation and the 2$p$ states of oxygen induced by the Jahn-Teller effect was proposed.\cite{jong} Usually, the weak Jahn-Teller effect occurs in Fe$^{3+}$ only in low-spin state. Further theoretical and experimental studies are required to correlate possible role of Jahn-Teller effect on the appearance of ferroelectricity in NFO. Finally, possible occurrence of ferroelectricity in other members of ferrite, AFe$_2$O$_4$ (A = Mn, Co, Cu, Zn) needs to be investigated for understanding origin of ferroelectric order and Verwey-type transition, which is underway. 

In conclusion, a FE order is observed for NFO at $\sim$ 98 K with a reasonably large value of the electric polarization of $\sim$ 0.29 $\mu$C/cm$^2$ for a 5 kV/cm poling field. A considerable magnetoelectric coupling is confirmed by a 17 \% change of the electric polarization upon application of 50 kOe magnetic field around 77 K, pointing a promising member of the type-II multiferroics. The FE transition is associated with a structural transition to a noncentrosymmetric $P4_{1}22$ structure from the inverse spinel structure, as confirmed by the analysis of the synchrotron diffraction studies. The polar order is found to be associated with an ordered occupancy of the Ni and Fe atoms at the octahedral sites of the $P4_{1}22$ structure. 
  
\vspace{0.2in}
\noindent
{\bf Acknowledgment}\\
S.G. acknowledges SERB, India project (Project No. SB/S2/CMP-029/2014) for the financial support. Portions of this work were carried out at the light source PETRA III of DESY, a member of the Helmholtz Association (HGF). Financial support (Proposal No. I-20170178) by the Department of Science \& Technology (Government of India) provided within the framework of the India@DESY collaboration is gratefully acknowledged. JKD and AC want to acknowledge DST-INSPIRE, India and UGC, India fellowships, respectively.

\end{document}